\documentstyle[aps,multicol,epsf]{revtex}
\begin{document}
\draft
\title{Stochastic Model for Surface Erosion Via
Ion-Sputtering: Dynamical Evolution from Ripple Morphology to
Rough Morphology}

\author{Rodolfo Cuerno$^1$, Hern\'an A. Makse$^1$, Silvina
Tomassone$^2$, Stephen T.
Harrington$^1$, and H. Eugene Stanley$^1$}

\address{$^1$ Center for Polymer Studies and Physics Department,
 Boston University, Boston, MA 02215}

\address{$^2$ Physics Department,
 Northeastern University, Boston, MA 02115}

\maketitle

\begin{abstract}
Surfaces eroded by ion-sputtering are sometimes observed to develop
morphologies which are either ripple (periodic), or rough
(non-periodic).  We introduce a discrete stochastic model that allows us
to interpret these experimental observations within a unified
framework. We find that a periodic ripple morphology characterizes the
initial stages of the evolution, whereas the surface displays
self-affine scaling in the later time regime.  Further, we argue that
the stochastic continuum equation describing the surface height is a
noisy version of the Kuramoto-Sivashinsky equation.
\end{abstract}




\begin{multicols}{2}

\narrowtext

A remarkable feature of erosion processes via ion sputtering in
amorphous materials is the formation of a pattern consisting of a ripple
structure, aligned in directions either parallel to or perpendicular to
that of the bombarding beam of ions \cite{revrough,revsput}.  Indeed, we
might expect that erosion tends to erase every possible
feature of the surface morphology, and that the presence of noise in the
system would further act against the formation of such a periodic
pattern.
Only recently have there been experimental \cite{chason}
and theoretical \cite{cb} attempts to understand the
formation of a ripple structure in the more general context of
non-equilibrium interface growth phenomena \cite{chason}. Many such
interfaces are
``rough'' and exhibit self-affine scaling at long distances and long
times \cite{revrough}; experimentally, one finds that surfaces
eroded by ion bombardment {\it also\/} exhibit self-affine scaling
behavior \cite{eklund}. An outstanding question is then how to reconcile
these observations with the formation of the periodic ripple structure.

In this Letter, we introduce a discrete stochastic model that
incorporates the main physical mechanisms believed to influence the
dynamics of the eroded surface morphology.  We show that this model is
characterized by an initial stage in which the surface morphology
displays a ripple structure, and that subsequent stages are
characterized by a crossover to a rough surface
in the universality class of the Kardar-Parisi-Zhang
(KPZ) equation \cite{kpz}.
We argue that the stochastic equation which provides the
continuum description of our model is a {\em noisy} version \cite{nks} of the
Kuramoto-Sivashinsky (KS) equation \cite{ks,crossover}.
Thus we interpret the formation
of a periodic pattern \cite{chason} and the development of rough
interfaces \cite{eklund} in the ion-sputtered systems as the early and
late regimes respectively of the same dynamical process.

We first consider the main mechanisms
\cite{neglect_redep} that determine the surface morphology
undergoing ion bombardment.

{\bf \it{(i)}} {\it Erosion} --- In ion sputtering, an initially flat
substrate is bombarded with a well-collimated beam of heavy ions
carrying a certain kinetic energy, and forming a precise angle with the
normal to the uneroded surface. The phenomena leading to
erosion take place within some finite distance from the surface. Namely,
the ions penetrate
inside the solid and induce along their path cascades of collisions among
the atoms of the substrate.  Atoms located at the surface may
be affected by these collisions and acquire enough energy to
leave the surface \cite{sigmund}.
Consequently, atoms located at the bottom of
troughs gain more energy on average and they are preferentially eroded as
compared to those on the peaks of crests \cite{bh}. This ``instability''
can be thought of as a {\it negative surface tension}, since the surface
tends to maximize its area.

{\it \bf(ii)} {\it Surface Diffusion} --- In physical systems there is a
stabilizing mechanism that balances the negative surface tension, {\it
surface diffusion}, which is always present at a non-zero temperature.
Particles on the surface tend to diffuse looking for highly coordinated
positions, a relevant phenomenon for sputtering \cite{chason,eklund} as
well as for molecular-beam epitaxy (MBE) \cite{mbe}.

When the angle between the local normal to the surface and the incident
beam approaches the grazing value, there is an increase in the
reflection of the ions by the surface and the rate of erosion
diminishes. This surface effect is beyond the approximations made in
Sigmund's theory of ion sputtering \cite{sigmund}, and is reflected in
the angle dependence of the sputtering yield, $Y(\varphi)$, defined to
be the number of eroded particles divided by the total number of
bombarding ions. Here $\varphi$ is the {\it local} angle of the ion
trajectories to the surface normal at each point \cite{function}.
Typically \cite{revsput}, $Y(\varphi)$ is symmetric around $\varphi=0$,
presents a maximum between 60$^{{\rm o}}$ and 80$^{{\rm o}}$,
and decreases to zero as $\varphi\to 90^{{\rm o}}$.

To define the model, we introduce two dynamical rules, one to account
for erosion and one to account for surface diffusion.  The
rule for erosion incorporates the unstable behavior described above as
well as the phenomenological dependence of the sputtering yield as a
function of $\varphi$.  The model for the case of
$1+1$ dimensions \cite{2+1} is defined on a square lattice of lateral
size $L$, with periodic boundary conditions in the horizontal direction.
The initial interface is a horizontal line separating occupied sites
(below) from empty sites (above).  We choose randomly a site $i$ at the
interface where $i=1,\ldots,L$.  The chosen site is subject to erosion
with probability $p$, or to diffusion with probability $1-p$, where the
rules are as follows:

{\it {\bf (i)}} {\it Erosion (probability p)---} \cite{normal} We compute
$\varphi\equiv\tan^{-1}[(h_{i+1}-h_{i-1})/2]$,
where $h_i$ is the height of the interface at site $i$, and apply the
erosion rule with probability $Y(\varphi)$, as given in Fig.\ $1a$. To
erode, we count the number of occupied neighbors inside a square
box of size $3 \times 3$ lattice spacings centered in the chosen site
$i$ (box rule). We empty the site with an erosion probability $p_e$
proportional to the number of occupied cells in the box (see Fig.\
$1b$). Thus the box rule favors the erosion of troughs as compared to
the peaks of crests, and therefore is the source of the instability in
the ion-sputtered system.

{\it \bf (ii)} {\it Surface Diffusion (probability $1-p$)---} A diffusive
move of the particle $i$ to a nearest neighbor column is attempted with
hopping probability
$w_{i \to f} \equiv \left[ 1 + \exp \left(\Delta
{\cal H}_{i \to f}/k_BT\right) \right]^{-1}$,
where $\Delta {\cal H}_{i \to f}$ is the energy difference between the
final and initial states of the move. Following \cite{ps}, we choose
${\cal H}\equiv (J/2) \sum_{\langle i,j \rangle} (h_i-h_j)^2$.

A continuum equation that describes the dynamics of the interface height
has the form
\begin{equation}
\label{fulleq}
\partial_t h( \mbox{\boldmath $x$}, t) \ = \ \nu \nabla^2 h -
\kappa{\nabla^4 h} + \eta(\mbox{\boldmath$x$}, t) + f_{Y}[h
(\mbox{\boldmath$x$}, t)].
\end{equation}
Here $h(\mbox{\boldmath$x$},t)$ is the height of the interface at
position \mbox{\boldmath$x$} and time $t$, $\nu$ is a negative surface
tension coefficient, $\kappa$ is a positive coefficient that accounts
for the surface diffusion, and $\eta(\mbox{\boldmath$x$},t)$ is a
Gaussian noise term with short range correlations and strength $2D$,
that accounts for the
fluctuations in the flux of incoming particles.  The functional
$f_{Y}[h]$ takes into account the contribution of nonlinear terms, which
appear in the equation of motion due to the effect of
$Y(\varphi)$, itself a nonlinear function of the local slope
$\nabla h \equiv \tan \varphi$. The nonlinearities
become more relevant in the equation of motion at the
late regime of the evolution
when large slopes
develop, see below.

First we study the model in the case $Y(\varphi)\equiv 1$, which
approximately holds at the early stages of evolution.  We focus on the
time dependence of the total interface width $W(t) \equiv \langle
L^{-1}\sum_{i=1}^L (h_i(t) - \overline{ h(t)} )^2
\rangle^{1/2}$, where
$\overline{h(t)}\equiv L^{-1} \sum_{i=1}^L h_i(t)$, and the brackets
denote an average over realizations of the noise.  The erosion rule
alone---corresponding to $p=1$, $Y(\varphi)\equiv 1$---leads to $W(t) \sim
t$, which can be attributed to $\nu$ in (\ref{fulleq}) being a negative
number \cite{aclar_expogrowth}. Moreover, by considering only the
erosion rule, but defined with probability $1-p_e$, we find that the interface
has the scaling properties of the Edwards-Wilkinson (EW) equation
\cite{ew} that is obtained from (\ref{fulleq}) with $\nu>0$, $\kappa=0$,
and $f_Y[h]=0$. We understand this result since one favors the erosion
of peaks as compared to valleys, leading to the smoothing mechanism
characteristic of a positive surface tension.  On the other hand, the
surface diffusion mechanism alone (for $p=0, Y(\varphi)\equiv1$) is well
described by (\ref{fulleq}) with $\nu=0$, $\kappa>0$, and $f_Y[h]=0$,
i.e., the linear MBE equation \cite{ps,mbe}.  When mechanisms {\it \bf (i)}
and {\it \bf(ii)} above are considered simultaneously for $J/k_B T=5$,
$p=0.5$, and $Y(\varphi)\equiv1$, we obtain the different stages of the
time evolution displayed in Fig.\ $2a$ \cite{other}.  There exists a
first region \cite{random} for which $W(t) \sim t^{\beta_1}$, with
$\beta_1 = 0.38\pm 0.02$, the growth exponent for the linear MBE
universality class, after which $W(t)$ scales with $\beta_2 > 0.5$
due to the instability caused by $\nu < 0$. In this case, a linear
stability analysis of (\ref{fulleq}) shows that there is a maximally
unstable mode in the system, $k_{m}= ( |\nu|/ 2\kappa)^{1/2}$, and
therefore the surface is almost periodic (inset in Fig.\ $2a$).

Next we consider the model with $Y(\varphi)$ shown in Fig.\ $1a$.  The
results are not expected to depend strongly on the specific form of
$Y(\varphi)$, so long as it preserves the existence of a maximum, and
$Y(0) \neq 0$, $Y(90^{{\rm o}})=0$\cite{carter2}.
Fig.\ \ref{sk} shows the structure factor $S(\mbox{\boldmath$k$})
\equiv \langle \widehat{h}(\mbox{\boldmath$k$},t) \widehat{h}(-
\mbox{\boldmath$k$},t) \rangle$ at the onset of the instability.  Here $
\widehat{h}(\mbox{\boldmath$k$},t)$ is the Fourier transform of $h_i(t)
- \overline{h(t)}$.  As we see, the early stages of the dynamics are
still dominated by the periodic ripple structure defined by the
competition between surface tension and surface
diffusion, described by the linear part of (\ref{fulleq}).

For later times, the large slopes built up by the instability induce
nonlinear effects, and the
interface results in a rough morphology (inset of Fig.\ $2b$).  In Fig.\
$2b$, we present the time evolution of $W(t)$ for the complete model.
We again observe a first regime \cite{random} with $\beta_1 = 0.38\pm
0.03$, followed by unstable erosion ($\beta_2 > 0.5$).  For later
stages, we find $\beta_3 = 0.23\pm 0.03$, consistent with EW, after
which a crossover to $\beta_4 = 0.28 \pm 0.03$ is found. Finally, the
width saturates due to the finite size of the system.
Note that the
value of the growth exponent for the KPZ equation is
$\beta_{\mbox{\scriptsize KPZ}} = 1/3$ \cite{kpz}. At saturation, $S(k)$
displays the small momenta behavior $ S(k,t) \sim k^{-2}$, consistent
with the scaling of both the EW and KPZ universality classes (see Fig.\ 3).
To determine if a KPZ nonlinearity is present in Eq.\ (\ref{fulleq}), we
compute the mean velocity $v(m)$ of the interface in the saturated
regime as a function of an average tilt $m\equiv\langle \nabla h\rangle$
imposed by using helical boundary conditions.  If we assume that
the relevant nonlinearity in (\ref{fulleq}) is of the KPZ type,
then $f_Y[h] = (\lambda/2) ({ \nabla h )}^2$. Taking spatial and noise
averages in (\ref{fulleq}), $v = v_{0} + (\lambda / 2)
m^2$, where $ v_{0}$ is the velocity of the untilted interface
\cite{krug}.  The parabolic shape of $v(m)$ obtained in our simulations
(see Fig.\ 4) leads to the conclusion that the long time and long distance
behavior of the model falls into the KPZ universality class. Moreover,
the continuum equation describing
the model ion-sputtered
surfaces is the {\em noisy} KS equation
\begin{equation}
\partial_{t} h = \nu \nabla^2 h - \kappa \nabla^4 h +
\frac{\lambda}{2} \left(\nabla
h\right)^2 + \eta(\mbox{\boldmath$x$},t) .
\label{nkseq}
\end{equation}

To compare
the {\em dynamics} of (\ref{nkseq}) with those obtained for the
discrete model, we integrate numerically Eq.\
(\ref{nkseq}) in $1+1$ dimensions. Fig.\ \ref{width-nks} shows the
behavior of the function $W(t)$. We observe the
same crossovers as in the model \cite{random}: $\beta_1 = 0.38\pm 0.03$
corresponding to the linear MBE case, followed by unstable growth
$\beta_2 > 0.5$. Then, a transition to EW behavior $\beta_3 = 0.25\pm
0.03$ is observed, after which the nonlinearities dominate and a KPZ
growth with $\beta_4 = 0.30\pm 0.03$ is obtained.
Finally, the
interface width saturates due to the finite size of the system.
Consistent with these numerical findings,
the late scaling of Eq.\ (\ref{nkseq}) has been shown through a
renormalization-group calculation \cite{nks}
to be that of the KPZ equation in 1+1 and 2+1 dimensions.
As we see, both in the model and the {\em noisy} KS, there is a long crossover
time from EW to KPZ behavior, responsible for the difference between
$\beta_4$ and $\beta_{KPZ}$, and for the narrow window in which $\beta_4$
is observed---we find that the width of this window increases systematically
with $L$. A similar phenomenon is well known to occur
in the {\em deterministic} KS equation in 1+1 dimensions, see Sneppen {\em
et al.} in \cite{crossover}.

Finally, we compare the results of the model with observations of recent
experiments \cite{2dmod}.
The experimental development of a ripple structure
\cite{chason} is well understood in terms of the unstable linear theory
of ion-sputtering describing the early stages of the time evolution of
the model presented here.  Moreover, the model predicts that in the late
regime the large slopes generated by the unstable growth trigger the
action of nonlinearities which stabilize the surface.  The nonlinearity
we find is of the KPZ type, consistent with the experimental observation
of KPZ scaling reported by Eklund {\it et al.}
\cite{eklund}.  To confirm the above picture, it would be of interest to
study experimentally if both regimes do effectively take place in the
time evolution of the {\it same} physical system.

We acknowledge discussions with A.\ L.\ Barab\'asi, G.\ Carter, S.\ Havlin,
and K.\ B.\ Lauritsen. R.\
C.\ acknowledges support from Ministerio
de Educaci\'on y Ciencia, Spain. The Center for Polymer Studies is funded by
NSF.

\begin{figure}
\caption{$(a)$ Sputtering yield $Y(\varphi)$ as a function of the
angle $\varphi$.
$(b)$ Box rule for erosion.
We define $p_e$ as the number of occupied neighboring sites
(grey squares)
inside the 3 $\times $ 3 box centered at site $i$ (black square),
normalized by 7.
The examples shown
correspond to $(i)$ $p_e=1$ and $(ii)$ $p_e=3/7$.}
\label{boxrule}
\end{figure}

\begin{figure}
\caption{$(a)$ Time evolution of the surface width for the cases
$Y(\varphi) \equiv 1$ and $L=50$.  The solid line is the consecutive
slope of the width, showing the value of the growth exponent $\beta$ in
each regime.  The inset shows the ripple structure of the interface
at $t=1000$.  The saturation observed in $W(t)$ is due
to the discreteness of the lattice: the erosion rule breaks down when
the local slopes of the interface are bigger than $3$ \protect\cite{kent}.
This effect can
be avoided by using a bigger box.  $(b)$ Interface
width as a function of time for the full model showing the regimes of
the evolution for $L=2048$.
As in $(a)$, the solid line is the consecutive slope.
The inset shows a portion of the rough morphology
at the late regime, where the self-affine scaling behavior
holds.  The arrows indicate the times at which the structure factor is
displayed in Fig.\ 3.}
\label{width-full}
\end{figure}

\begin{figure}
\caption{Structure factor computed
using the full model for a system with $L=2048$.  For $t=300$, averaged
over 2600 noise realizations ($\diamond$), and for $t=1.7 \times 10^6$,
averaged over 39 realizations ($\bullet$) (see arrows in Fig.\ 2b).  The
solid line is a fit to the exact solution of the discretized
linear part of Eq. $(1)$.
The dashed straight line has slope $-2$.}
\label{sk}
\end{figure}

\begin{figure}
\caption{Plot of the mean velocity $v(m)$
as a function of the average tilt $m$ of the interface,
calculated in the saturated regime for
$L=128$, and averaged over $810$ noise realizations.
The solid line is a fit to a parabola.}
\label{parabola}
\end{figure}

\begin{figure}
\caption{Log-log plot of the interface width as a function of
time obtained in the numerical integration of Eq.\
$(2)$. The mesh is $\Delta x=1.0$ and the time step $\Delta t=0.005$.
Parameter values are $\nu=-1$, $\kappa=1$, $\lambda=4$, $D=0.5$, and
$L=1024$.  The solid line gives the consecutive slopes.}
\label{width-nks}
\end{figure}

\end{multicols}


\begin{references}

\bibitem{revrough}
P. Meakin, Phys. Rep. {\bf 235}, 189 (1993);
T. Halpin-Healey and Y.-C.  Zhang, {\it ibid.}
{\bf 254}, 215 (1995); A.-L. Barab\'asi and H. E. Stanley, {\em Fractal
Concepts in Surface Growth} (Cambridge University Press, Cambridge
1995)

\bibitem{revsput}
G. Carter {\it et al.},
in {\em Sputtering by Particle Bombardment}, Vol. II,
edited by R. Behrisch,
(Springer-Verlag, Heidelberg 1983), p. 231.

\bibitem{chason}
E. Chason {\it et al.},
Phys. Rev. Lett. {\bf 72}, 3040 (1994); T. M. Mayer {\it et al.},
J. Appl. Phys.  {\bf 76}, 1633 (1994).

\bibitem{cb}
R.\ Cuerno and A.\ L.\ Barab\'asi, Phys. Rev. Lett. {\bf 74}, 4746 (1995).

\bibitem{eklund}
E. A. Eklund {\it et al.},  Phys. Rev. Lett.
{\bf 67}, 1759 (1991);
E. A. Eklund  {\it et al.},
Surf. Sci. {\bf 285}, 157 (1993); J. Krim  {\it et al.},  Phys. Rev. Lett.
{\bf 70}, 57 (1993); H.-N. Yang, G.-C. Wang, and T.-M. Lu,
Phys. Rev. {\bf B 50}, 7635 (1994).

\bibitem{kpz}
M. Kardar, G. Parisi and Y.-C. Zhang, Phys. Rev. Lett. {\bf 56}, 889
(1986).

\bibitem{nks}
R. Cuerno and K. Lauritsen, Phys. Rev. {\bf E} {\bf 52}, 4853 (1995).
For $d>2$, see L. Golubovi\'c and R. Bruinsma, Phys. Rev. Lett.  {\bf 66},
321 (1991); {\em ibid.} {\bf 67}, 2747 (E) (1991).

\bibitem{ks}
Y.\ Kuramoto and T.\ Tsuzuki, Prog.\ Theor.\ Phys.\ {\bf 55}, 356
(1977); G.\ I.\ Sivashinsky, Acta Astronaut. {\bf 6}, 569 (1979).

\bibitem{crossover}
S. Zaleski, Physica {\bf D 34}, 427 (1989);
K. Sneppen {\it et  al.}  Phys. Rev.
{\bf A 46}, R7351 (1992); F. Hayot {\it et al.},
Phys. Rev. {\bf E 47}, 911 (1993).

\bibitem{neglect_redep}
Here we neglect redeposition of the eroded material, as well as
shadowing effects among different surface features. These assumptions have
been shown \cite{chason}
to hold at the initial stages of surface evolution. The roughening
taking place in the model at late time stages is again consistent with them.

\bibitem{sigmund}
P. Sigmund, Phys. Rev. {\bf 184}, 383 (1969).

\bibitem{bh}
P. Sigmund, J. Mat. Sci. {\bf 8}, 1545
(1973); R. M. Bradley and J. M. E. Harper,
J. Vac. Sci. Technol. {\bf A 6}, 2390
(1988).

\bibitem{mbe}
C. Herring, J. Appl. Phys. {\bf 21}, 301 (1950); W. W. Mullins, J. Appl. Phys.
{\bf 28}, 333 (1957); in the context of stochastic models, see
D. E. Wolf and J. Villain, Europhys. Lett. {\bf 13}, 389 (1990);
S. Das Sarma and P. I. Tamborenea, Phys. Rev. Lett. {\bf
66}, 325 (1991).

\bibitem{function}
When we fix the
type of
bombarding ion, the material of the substrate,
and the energy of the incoming beam,
$Y(\varphi)$ is a function of $\varphi$ only.

\bibitem{2+1}
The generalization to $2+1$ dimensions will be reported elsewhere.

\bibitem{normal}
We consider
normal incidence of the ions onto the substrate.

\bibitem{ps}
M. Siegert and M. Plischke, Phys. Rev. {\bf E 50}, 917 (1994).

\bibitem{aclar_expogrowth}
Strictly speaking, $\nu < 0$ implies exponential growth for $W$,
which can not be attained in the present definition of the box rule,
since at most only $L$ particles can be eroded per unit time, leading
to $W \sim t $ as the fastest growth for the width.

\bibitem{ew}
S. F. Edwards and D. R. Wilkinson, Proc. R. Soc. Lond. {\bf A 381}, 17
(1982).

\bibitem{kent}
K. Lauritsen (private communication).

\bibitem{other}
Qualitatively the results do not change when other
values of $J/k_B T$ and $p$
are considered.

\bibitem{random}
An initial random erosion regime ($\beta_0=0.5)$
is also observed before correlations build up in the system.

\bibitem{carter2}
G. Carter {\it et al.}, Surf. Interface Anal. {\bf 20}, 90 (1993);
A. N. Protsenko, Nucl. Instr. and Meth. {\bf B 82}, 417 (1993).

\bibitem{krug}
J. Krug and H. Spohn, Phys. Rev. Lett. {\bf 64}, 2332 (1990).

\bibitem{2dmod}
Based on our identification of Eq. (\ref{nkseq}) as describing the model,
and the scaling of this equation in 2+1 dimensions predicted in
\cite{nks}, we expect
our results
to carry over for the physical case of two-dimensional surfaces.
An important ingredient in 2+1 dimensions is the anisotropy between the
two substrate directions induced by the ion beam. This
leads to
effects such as the change in the
orientation of the ripples as a function of the angle of incidence
\cite{bh}, and could be
accounted for by introducing a non-zero angle of incidence.

\end{references}
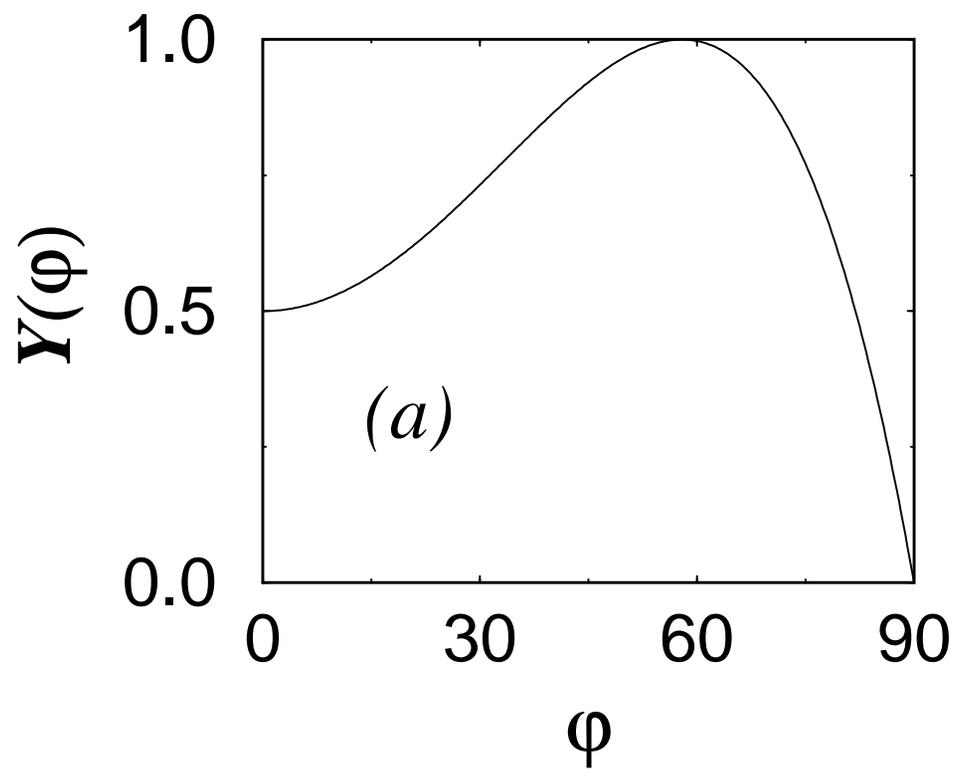
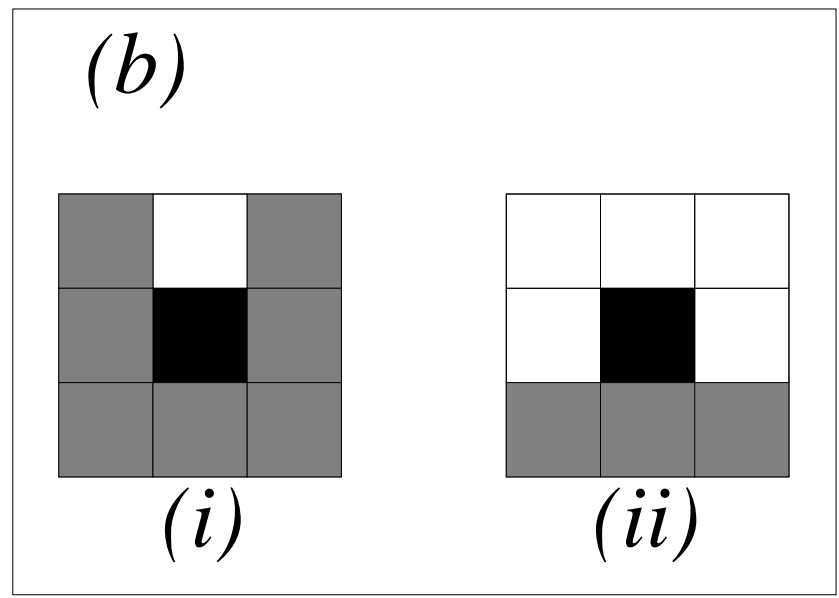
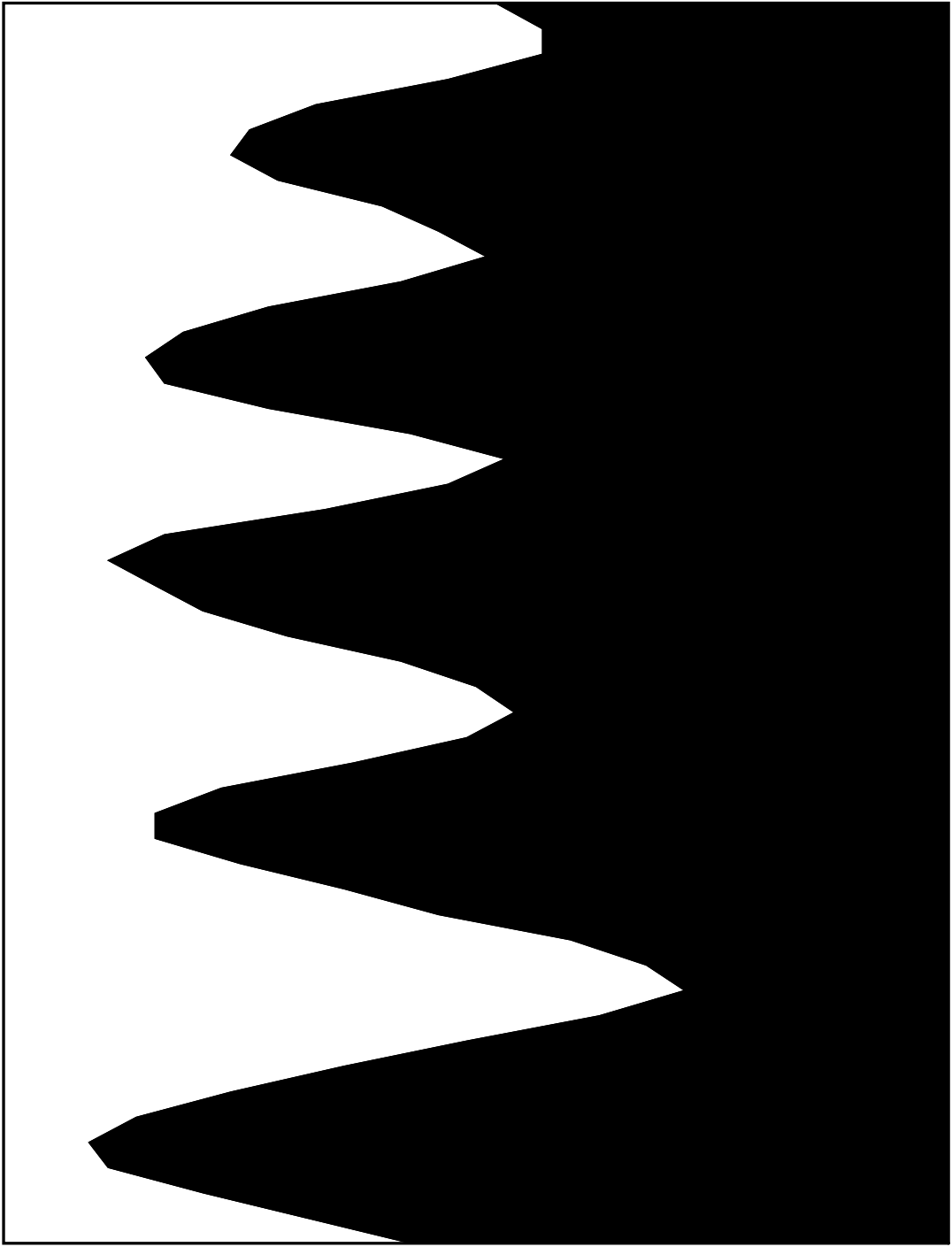
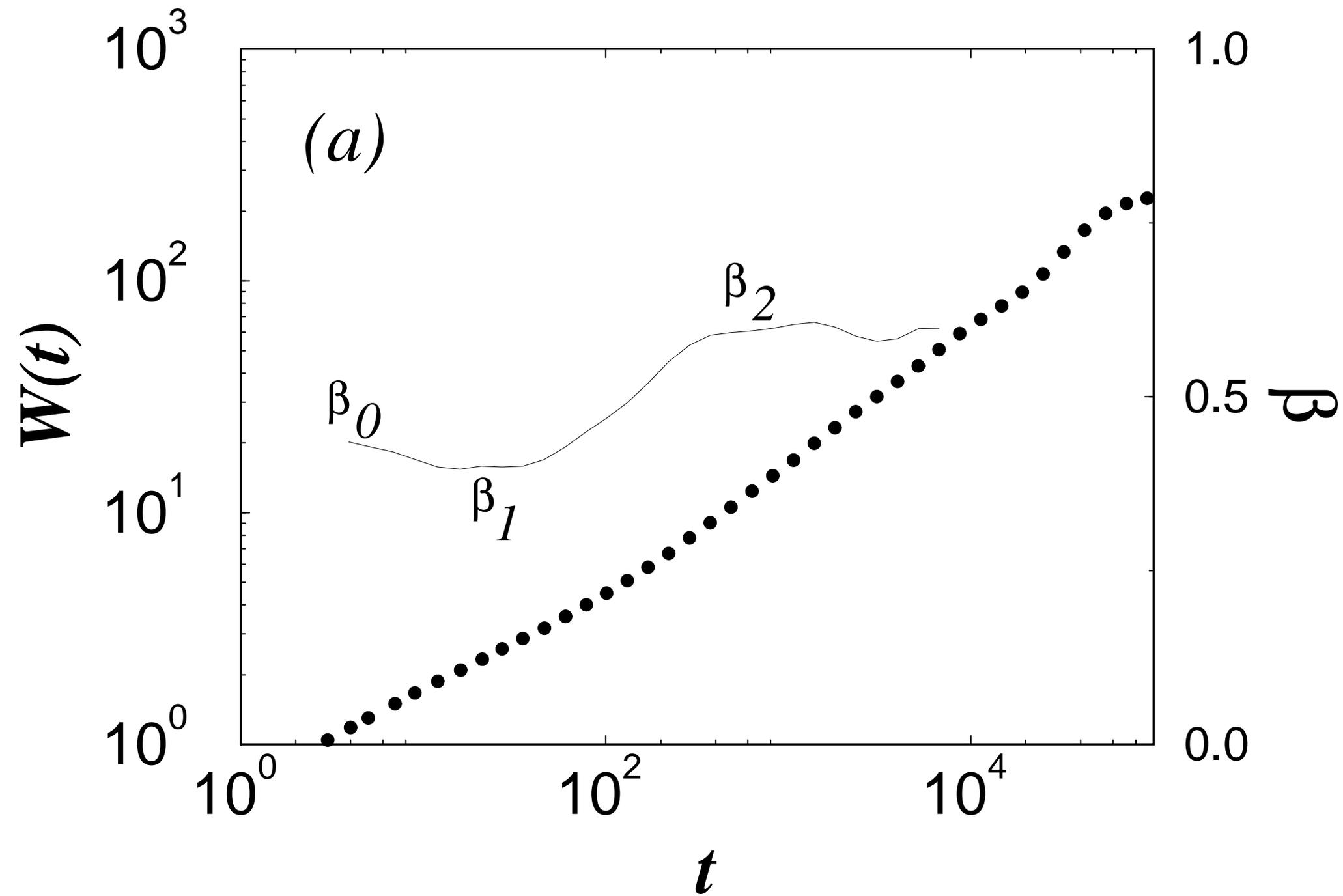
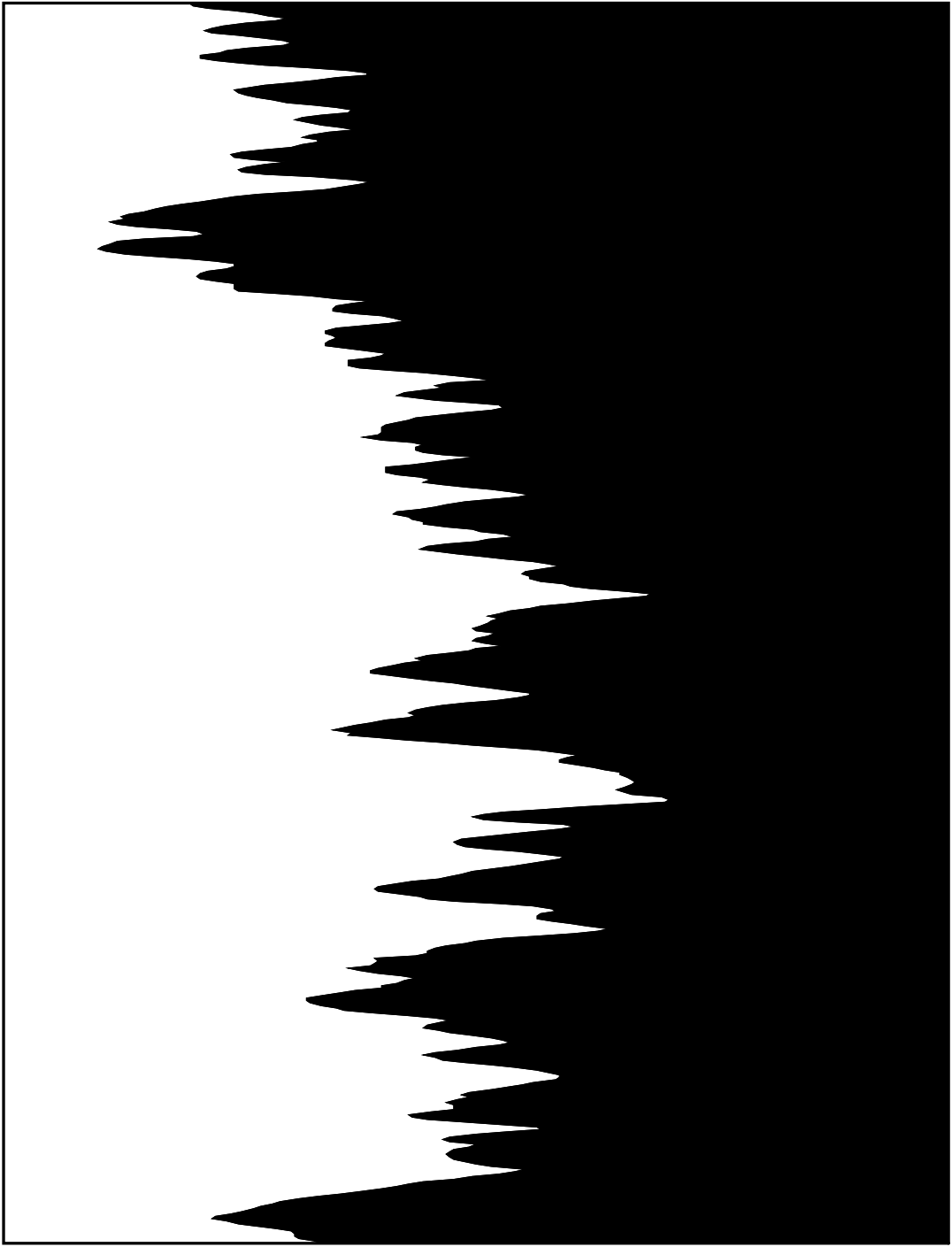
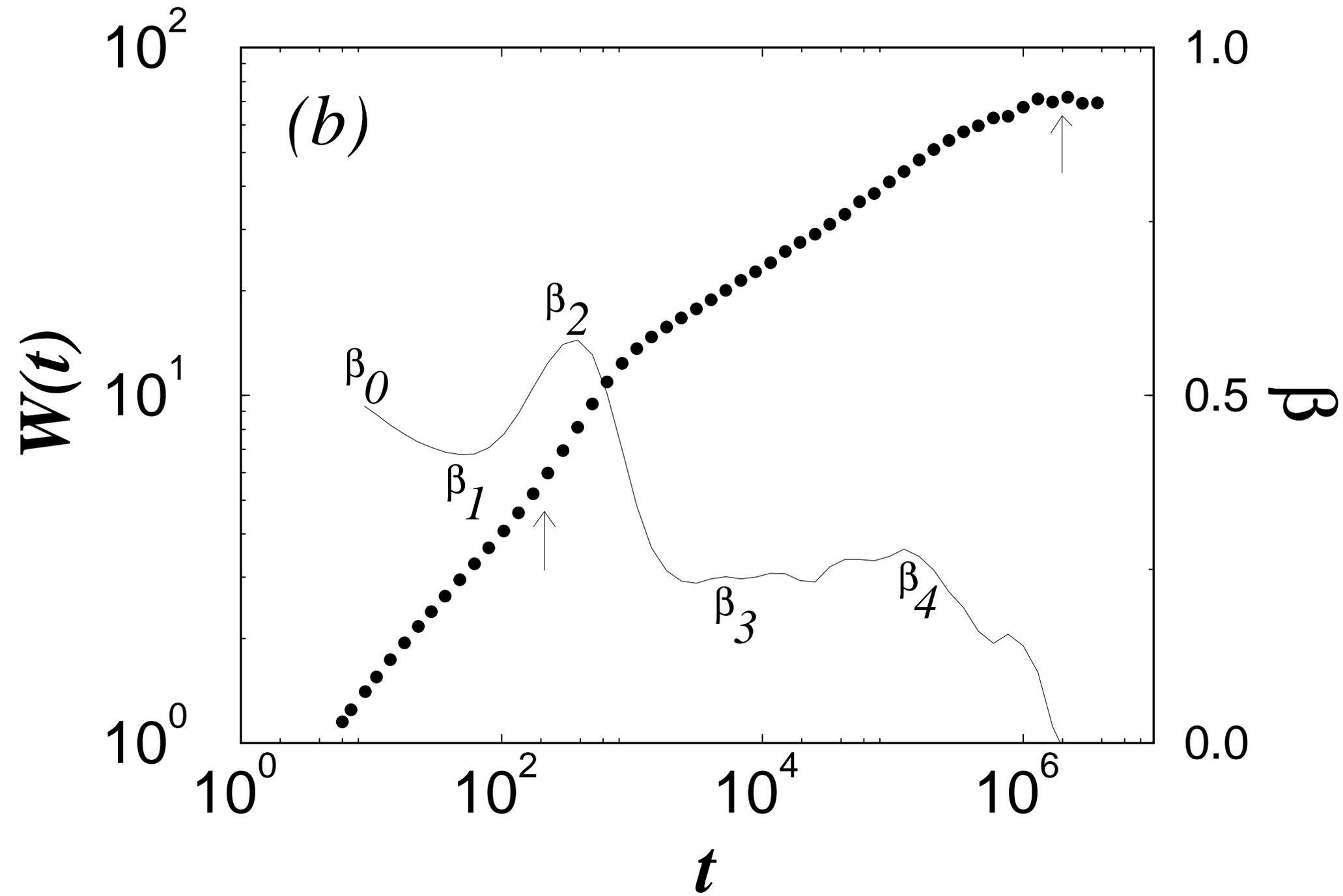
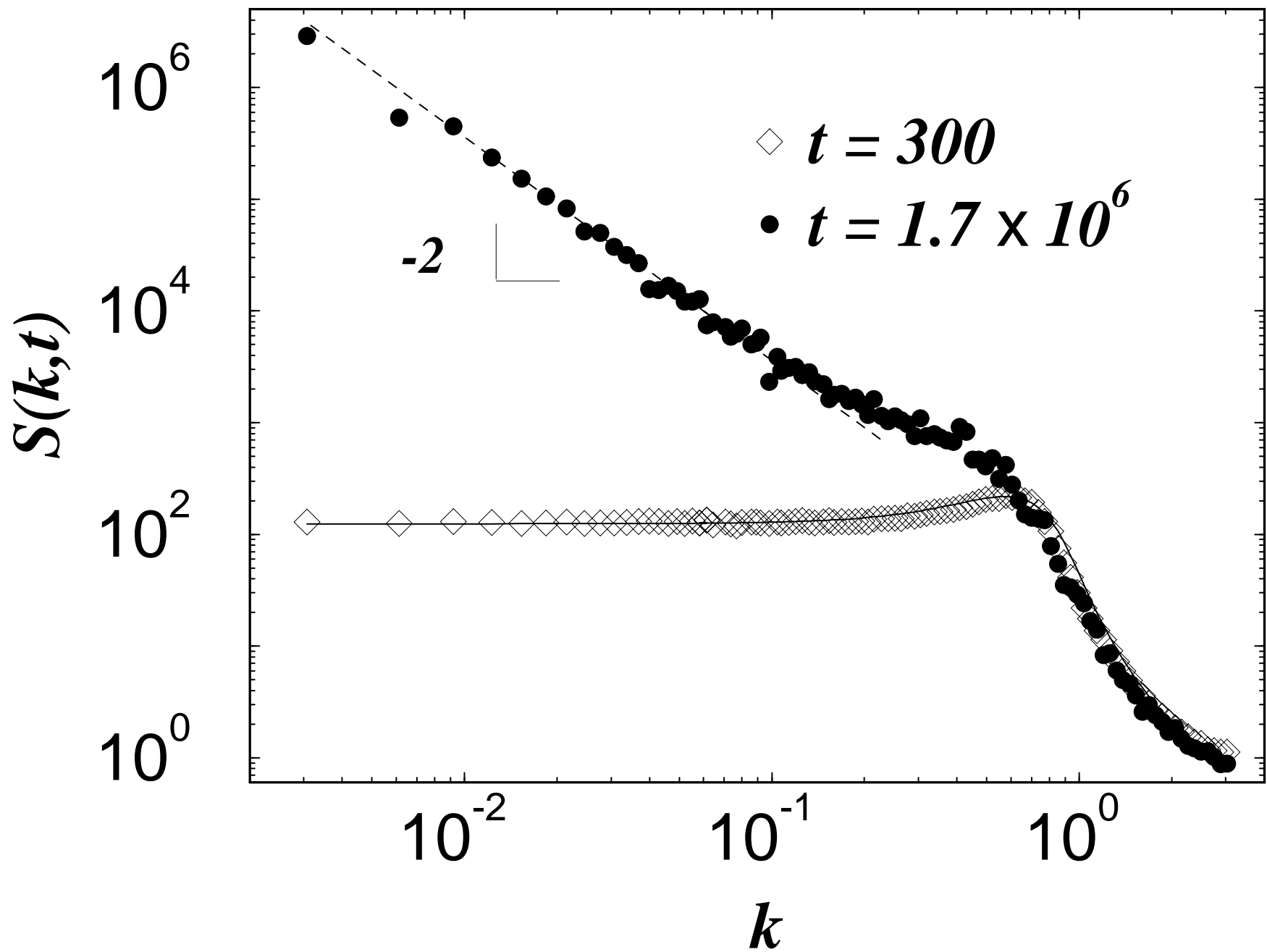
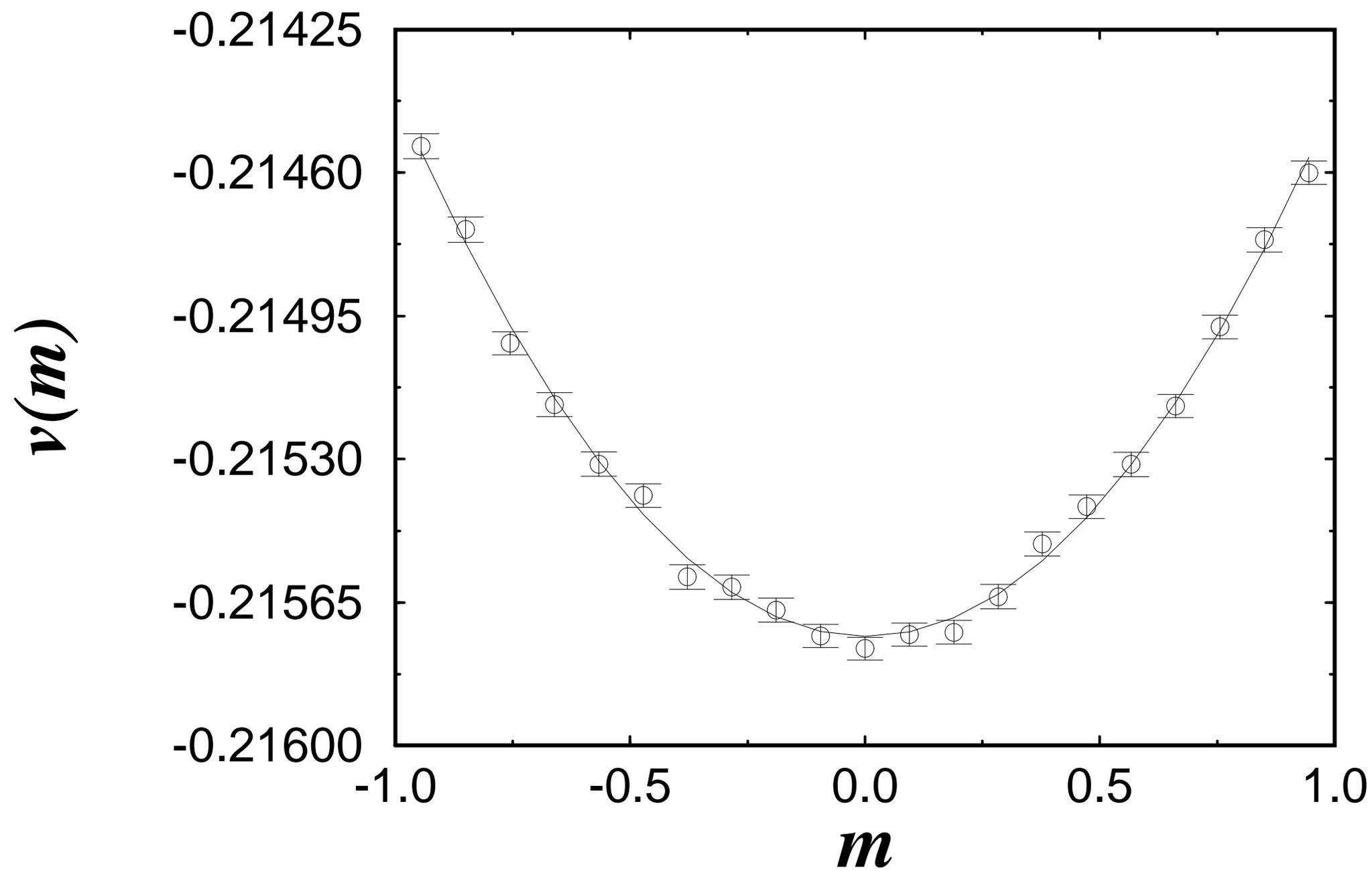
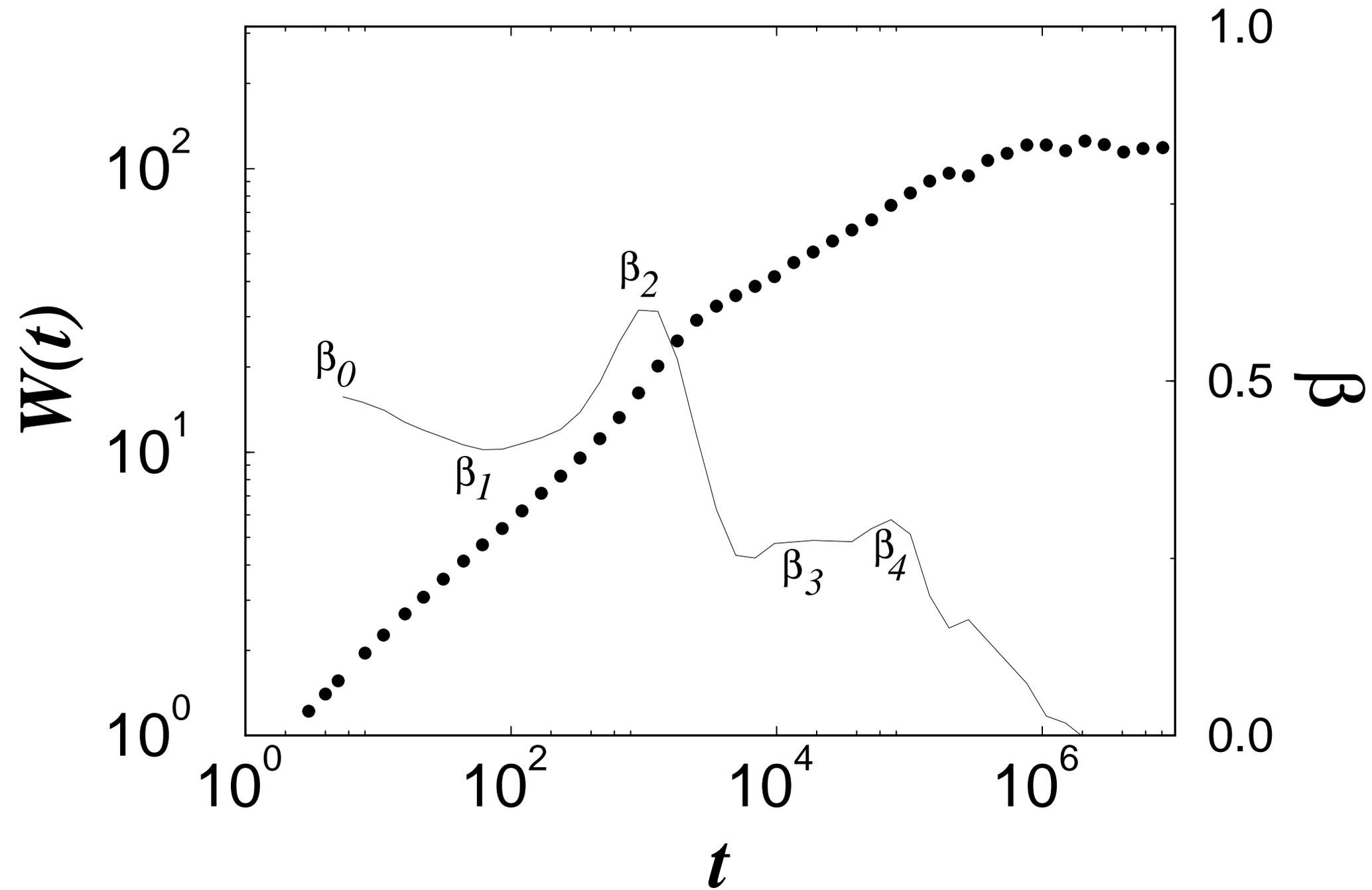
\end{document}